\documentstyle[preprint,aps,epsf]{revtex}
\tightenlines
\begin{document}
\draft

\title{
\vspace{-3.0cm}
\begin{flushright}  
{\normalsize UTHEP-376}\\
\vspace{-0.3cm}
{\normalsize UT-CCP-32}\\
\vspace{-0.3cm}
{\normalsize MPI-PhT/97-77}\\
\vspace{-0.3cm}
{\normalsize December 1997 }\\
\end{flushright}
%\vspace*{2.0cm}
%
The domain-wall model in an asymptotic-free dynamics
}

\author{ S. Aoki $^{a,b}$\footnote{aoki@mppmu.mpg.de} 
and K. Nagai $^c$\footnote{nagai@rccp.tsukuba.ac.jp}}
\address{$^a$ Institute of Physics,  University of Tsukuba,
         Tsukuba, Ibaraki 305, Japan}
\address{$^b$ Max-Planck-Institute f\"ur Physik,
         F\"ohringer Ring 6, D-80805 M\"unchen, Germany}
\address{$^c$ Center for Computational Physics,  University of Tsukuba,
         Tsukuba, Ibaraki 305, Japan}

\date{\today}
\maketitle

\begin{abstract}
We investigate a possibility that the rough gauge problem,
which have appeared to be a main reason for failures
of lattice chiral gauge theories,
is cured by an asymptotic-free dynamics.
Taking the domain-wall model in 2(+1) dimensions with SU(2) gauge group,
we carry out the quenched simulation of gauge fields in the extra dimension.
By studying fermion spectra in several volumes,
we show that
the chiral zero modes exist on the wall
without having the spontaneous symmetry breaking
thanks to the asymptotic-free dynamics.
This result suggests that the rough gauge problem
is solved in some class of lattice chiral fermions as well as
in 4 dimensions if an asymptotic-free dynamics is incorporated.

\end{abstract}

\narrowtext
\section{Introduction}
\label{sec:intro}
A definition of lattice chiral gauge theories is one 
of the long-standing problems in the lattice field theory,
in spite of its success as the non-perturbative calculational
method for vector gauge theories such as QCD.
The difficulty is related to the no-go theorem\cite{nielnino},
which proves, under the mild conditions, that
it is impossible to define chirally invariant lattice fermion
without species doubling.
Despite many attempts to put chiral gauge theories on the lattice
for many years,
so far none of them has been proven to work successfully.

Through analyses of many unsuccessful 
attempts\cite{kaplan,wilyu,aokinagai,waveg,latchiral}, however,
a common reason of failures has been recently recognized, 
called the ``rough gauge problem''\cite{latchiral}, 
which we will explain below.
A proposed lattice fermion action for one fermion field (in an anomalous 
representation) necessarily breaks gauge invariance,
to reproduce the chiral gauge anomaly in the weak gauge 
coupling limit,
though it may be cancelled by the contribution from other fermion fields.
Because of this explicit gauge breaking,
the gauge degree of freedom,
which varies independently at each site,
cannot be gauged away and thus interacts with the fermion,
so that the fermion is not a free field
even in the weak gauge coupling limit.
It is now widely accepted that 
the roughness of the gauge degree of freedom 
is an essential reason for failure of the proposals 
such as the Wilson-Yukawa model\cite{wilyu},
the U(1) original domain-wall model\cite{aokinagai}
and its variant\cite{waveg}, 
though the detail of the failure depends on the proposal.

By adding the kinetic term for the gauge degree of freedom to the action,
one can make it smooth enough to have a desired chiral fermion spectrum.
It has been shown in 4 dimensional models that
the dynamics of this smoothed gauge degree( = scalar field)
simultaneously leads to the condensation of the scalar field,
which makes the gauge field massive in lattice unit,
so that no chiral gauge theory is defined in the continuum limit.
This unsatisfactory correspondence that symmetric phase - vector spectrum or
Higgs phase - chiral spectrum seems to be established almost for all models.

Recently, it is, however, claimed that 
this correspondence is not true and thus the rough gauge problem might be 
solved if the dynamics of the gauge degree of freedom becomes 
asymptotic-free\cite{gdegree}.
It is actually shown in the spin wave calculation
that the rough configuration decouples from the fermion
in the overlap formula\cite{overlap}.

In this paper we pursue this possibility further,
using the quenched numerical simulation of the original domain-wall model,
where the gauge degree of freedom corresponds to the gauge field in the extra
dimension.
Previously we studied
the original domain-wall model with U(1) gauge group both in 2(+1) and 4(+1)
dimensions, whose gauge field in the extra dimension has
the asymptotic {\it non}-free dynamics and indeed found
no chiral zero mode in the symmetric phase.
In contrast,
we consider here the SU(2) gauge group in 2(+1) dimensions,
where the gauge field in the extra dimension is equivalent to
an asymptotic-free SU(2) $\times$ SU(2) non-linear $\sigma$ model.
Therefore there is only one phase, symmetric phase, in the model.
By the quenched Monte-Carlo simulation of the model
we investigate an existence of chiral zero modes
near and below the scaling region of the model.
This paper is organized as follows.
We explain the original domain-wall model
in Sec.\ref{sec:model}.
In Sec.\ref{sec:analysis}
the method of our analyses is explained, then
the results are shown.
Our conclusion and discussion are given in
Sec.\ref{sec:conclusion}.

\section{the domain-wall model}
\label{sec:model}
The original domain-wall model is 
the $2k(+1)$-dimensional Wilson fermion
whose mass term has a kink-like shape in the extra dimension,
vectorically interacting with the $2k(+1)$-dimensional gauge fields.
The action of the original model,
reformulated in Ref.\cite{naraneu}
in terms of a $2k$-dimensional theory,
are given as 
\begin{equation}
S = S_{G} + S_{F} , 
\end{equation}
where $S_{G}$ is the gauge action and
$S_{F}$ is the fermionic action.
$S_{G}$ is given by
\begin{equation}
S_{G} = \beta \sum_{n,\mu > \nu} \sum_{s} {\rm Re Tr} \left[ 
U_{\mu \nu}^{P}(n,s) \right]
+ \beta_{s} \sum_{n,\mu} \sum_{s} {\rm Re Tr} \left[ 
U_{\mu D}^{P}(n,s) \right]  , 
\end{equation}
where $\mu ,  \nu$ run from $1$ to $2 k$, 
$n$ is a $2 k$-dimensional lattice point, 
and $s$ is a coordinate of an extra dimension.
$U_{\mu \nu}^{P}(n,s)$ is a $2 k$-dimensional plaquette
and $U_{\mu D}^{P}(n,s)$ is a plaquette
containing two link variables in the extra direction
($D=2k+1$).
$\beta$ is the inverse gauge coupling for the plaquette 
$U_{\mu \nu}^{P}$ 
and $\beta_{s}$ is the one for the plaquette $U_{\mu D}^{P}$ .
In general, we can take $\beta \neq \beta_{s}$.
The fermionic action $S_{F}$ on the Euclidean lattice
is given by
\widetext
\begin{eqnarray}
 S_{F}& = & \frac{1}{2} \sum_{n \mu} \sum_s 
    \bar{\psi}(n, s) \gamma_\mu \left[ U_{\mu}(n, s)\psi(n + \mu, s)
    - U_{\mu}^{\dag}(n - \mu, s) \psi(n - \mu, s) \right]  \nonumber \\
& & \mbox{}+ \sum_n \sum_{s,t} \bar{\psi}(n, s) \left[ M_0 P_R 
     + M_0^{\dag} P_L \right]_{s, t} \psi(n, t)  \nonumber \\
& & \mbox{}+ \frac{1}{2} \sum_{n \mu} \sum_s
        \bar{\psi}(n, s) \left[ U_{\mu}(n, s) \psi(n + \mu, s)
    + U_{\mu}^{\dag}(n - \mu, s) \psi(n - \mu, s) -2\psi(n, s) \right],
\label{eqn:fermion} 
\end{eqnarray}
\narrowtext
where $s , t$ are an extra coordinates , 
$P_{R/L} = \frac{1}{2} (1 \pm \gamma_{2k+1})$ , 
\begin{equation}
\left\{ 
\begin{array}{l}
 ( M_0 )_{s,t} = U_{D}(n, s) \delta_{s + 1 , t} - a(s) \delta_{s,t} \\
 ( M_0^{\dag} )_{s,t} = U_{D}^{\dag}(n, s-1) \delta_{s - 1 , t} - a(s) 
\delta_{s,t} . 
\end{array}
\right.
\end{equation}
Here $U_{\mu}(n, s) , U_{D}(n, s)$  are link variables 
connecting a site $(n,s)$ to $(n+\mu,s)$ or $(n,s+1)$, respectively.
Because of a periodic boundary condition in the extra dimension, 
$s , t$ run from $-L_{s}$ to $L_{s} - 1$ , 
and $a(s)$ is given by 
\begin{eqnarray}
  a(s) & = & 1 - m_0 \, {\rm{sign}}
\left[( s + \frac{1}{2} ) \, {\rm{sign}}( L_{s} - s - \frac{1}{2} ) \right] \nonumber \\
 & = & \left\{
\begin{array}{ll}
1 - m_0  & \mbox{}  ( - \frac{1}{2} < s < L_{s} - \frac{1}{2} ) \\
1 + m_0  & \mbox{}  ( - L_{s} - \frac{1}{2} < s < - \frac{1}{2} ) \, ,
\end{array}
\right.
\end{eqnarray}
where $m_{0}$ is the height of the domain wall mass.
It is easy to check 
that the above fermionic action is identical to the one
in $2k(+1)$ dimensions, proposed by Kaplan\cite{kaplan,naraneu}.
This model describes chiral fermions for smooth back-ground gauge fields
in perturbation theory\cite{perturbation}.
We studied the weak coupling limit,
where the physical gauge coupling constant 
$g \rightarrow 0$.
In this case,
all gauge fields in the physical dimensions can be gauged away, 
while the gauge field in the extra dimension is still dynamical 
and its dynamics is controlled by $\beta_s$. 
Instead of the gauge degrees of freedom
in the edge of the waveguide
$(2k+1)$th component of gauge link variables represent 
roughness of $2k$-dimensional gauge fields.
An important question is whether the chiral zero mode on the domain wall
survives in the presence of this rough dynamics.

In this weak coupling limit,
the dynamics of the gauge field is equivalent to
$2k$-dimensional scalar model with $2 L_s$ independent copies 
where $2 L_s$ is the number of sites in the extra dimension,
as following way:
\begin{equation}
S_{G} = \beta_{s} \sum_{s} \sum_{n,\mu} 
\left\{ 1 - {\rm Re Tr} \left[ V(n,s) V^{\dag}(n+\mu,s) \right] \right\}, 
\label{eqn:gauge}
\end{equation}
where $V(n,s)=U_D(n,s)$.
If the gauge group is SU(N), the above action is equivalent to 
an SU(N) $\times$ SU(N) non-linear $\sigma$ model at each $s$,
therefore it is asymptotic-free in 2 dimensions and
has only a symmetric phase  for all $\beta_s$.
Larger $\beta_s$ smoother $V(n,s)$ without having non-zero condensation of
$V(n,s)$.
We can increase $\beta_s$ as much as we want till
the chiral zero modes will appear on the wall.
If it appears at finite $\beta_s$ we can conclude that the
chiral fermion exists in the symmetric phase, so the model passes the first
test toward the construction of lattice chiral gauge theories.
In this paper we take SU(2) as the gauge group and
investigate it in $g\rightarrow 0$ limit by the quenched simulation.

\section{analyses and results}
\label{sec:analysis}
\begin{figure}
\centerline{\epsfxsize=15cm \epsfbox{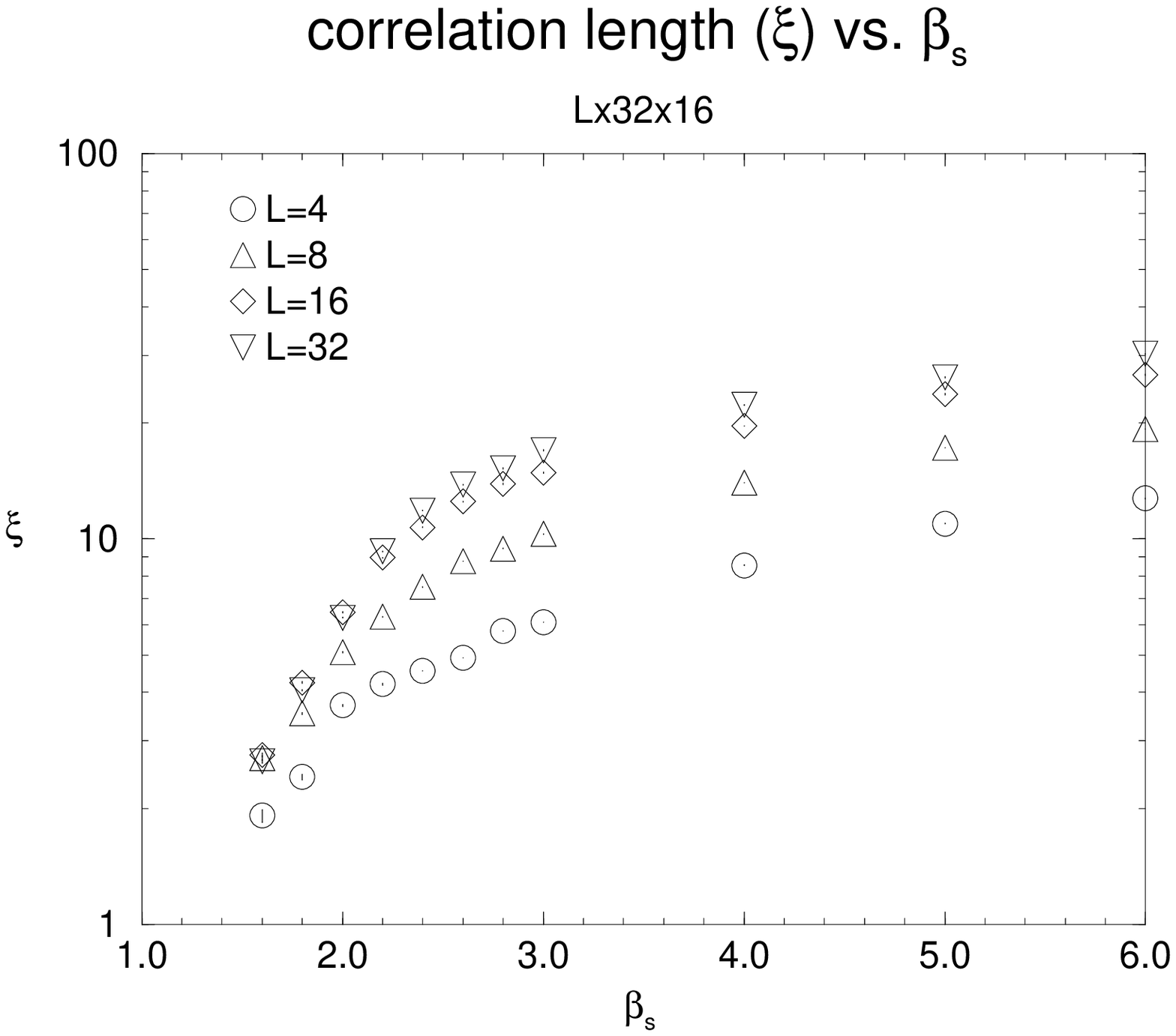}}
\caption{
Correlation length ($\xi$) vs. $\beta_{s}$
on a $L \times 32 \times 16$ lattices 
with $L=$4(circles) , 8(up triangles) 
, 16(diamonds) and 32(down triangles).
The error bar is smaller than the symbol.
}
\label{fig:corr}
\end{figure}

The method of simulation and analysis is essentially the same in 
Ref.\cite{aokinagai} except the gauge group.
In order to determine the choices of $\beta_s$ 
for the fermion propagator 
calculation, we first calculate the correlation length of the scalar field
$V(n,s)$ and the result is given in Fig.~\ref{fig:corr}
as a function of $\beta_s$.
Form this figure we estimate that 
the scaling region, where the correlation length grows
exponentially in $\beta_s$, begins around $\beta_s=2.0$.
This estimation is consistent with the one estimated from 
the susceptibility\footnote{Note
that the notation of the coupling constant in their study
is twice smaller than ours.}\cite{WZW}.
We also see from the data between $L=16$ and 32
that the finite size effect in the correlation length is 
negligible at $\beta_s \leq 2.0$ for $L=16\equiv L_{min}$.
The relation expected from consistency that $\xi < 8 =L_{min}/2 $
also holds at $\beta_s \le 2$.

 From the above result on the correlation length
we decide to calculated the fermion propagators
at $\beta_s$ = 1.0 and 2.0 for several values of $m_0$ on 
different sizes of lattices, using a quenched approximation.
The fermion mass in the 2-dimensional theory is extracted
from the fermion propagator within the $s=0$ layer
as follows\footnote{The details
of this method and the notation are 
explained in Ref.~\cite{aokinagai}. 
However, note that $G_L$$(G_R)$ here is denoted
$G_R$$(G_L)$ in Ref.~\cite{aokinagai} }.
The fermion propagator $G(p)_{st}$ in the 2 dimensional momentum space and
the coordinate of the extra dimension is written as
\begin{eqnarray}
G(p)_{s,t} &=& 
\left[ \left( 
        - i \sum_\mu \gamma_\mu \bar{p}_\mu + M \right) 
G_R(p) \right]_{s,t} P_L 
+   \left[ \left(
         - i \sum_\mu \gamma_\mu \bar{p}_\mu + M^{\dag} \right) 
G_L(p) \right]_{s,t} P_R .
\label{eqn:propagator}
\end{eqnarray}
Extracting $G_L(p)_{s=t=0}$ and $G_R(p)_{s=t=0}$ and extrapolate
them to $p=0$ we obtain the (effective) fermion mass $m_f$ as
\narrowtext
\begin{eqnarray}
(m_f^L)^2 &=& \lim_{p\rightarrow 0} \frac{1}{G_L(p)_{s=t=0}}\nonumber \\
(m_f^R)^2 &=& \lim_{p\rightarrow 0} \frac{1}{G_R(p)_{s=t=0}}
\end{eqnarray}
where $L(R)$ for $m_f$ stands for the left-handed(right-handed).
We take a periodic boundary condition 
in the first- and the extra-directions
and anti-periodic boundary in the second-direction
to avoid possible singularity of a massless fermion.
The fermion propagator is obtained from the average over
50 configurations 
and the errors are estimated by jack-knife method with an unit bin size.

\begin{figure}
\centerline{\epsfxsize=12cm \epsfbox{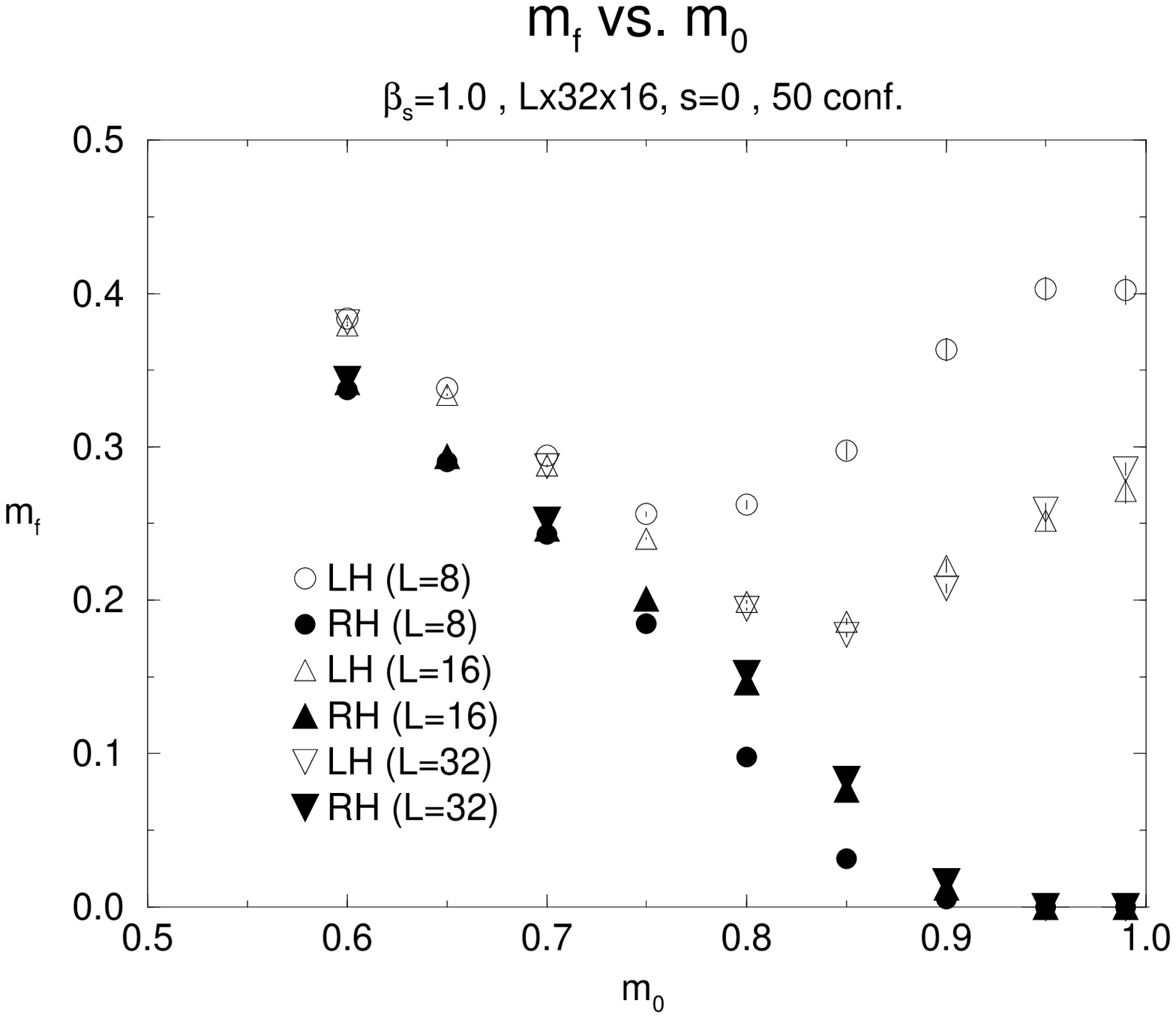}}
\centerline{\epsfxsize=12cm \epsfbox{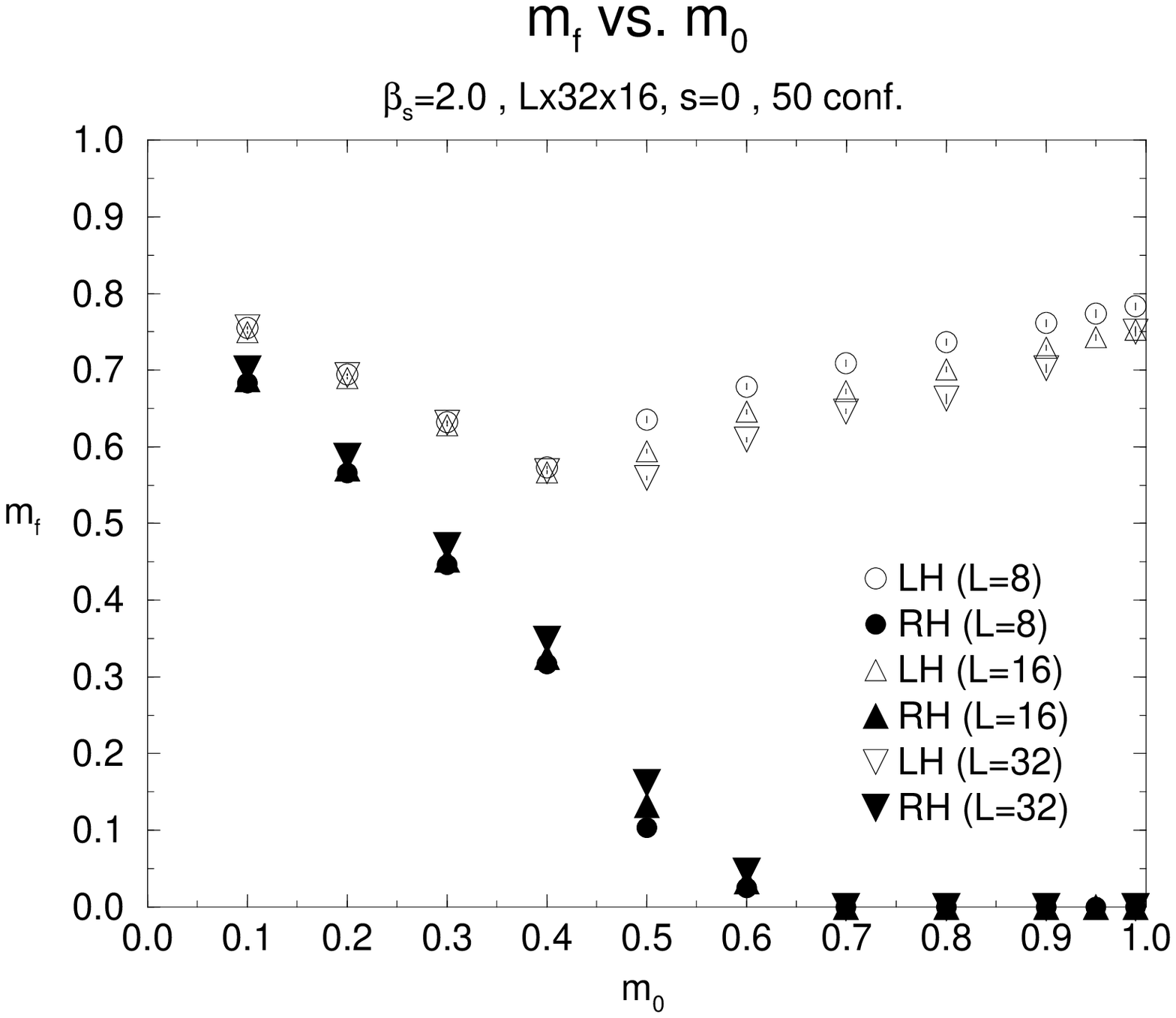}}
\caption{
$m_{f}$ vs. $m_{0}$ 
at $\beta_{s} = $1.0 and 2.0 
on a $L \times 32 \times 16$ lattices 
with $L=$8(circles) , 16(up triangles) and 32(down triangles)
obtained form the fermion propagator on the domain wall at $s=0$,
for the right-handed fermion (solid symbols) 
and the left-handed fermion (open symbols).
}
\label{fig:mfb}
\end{figure}

In Fig.~\ref{fig:mfb} 
the fermion mass extracted in this way is plotted
as a function of $m_0$ at $\beta_s=$1.0 and 2.0 
on $L \times 32 \times 2 L_s$ lattices with $L=$8, 16, 32 and $L_s = 8$.
This figure shows
that  the right-handed mode becomes massless
when $m_0$  is grater than some critical value: $m_0^c$.
For example, $m_0^c \sim 0.8 - 0.95$ at $\beta_s=1.0$
and $m_0^c \sim 0.4 - 0.6$ at $\beta_s=2.0$.
On the other hand,
the left-handed mode are massive at all $m_0$.
It also shows that
the finite size effect is small between $L=16$ and $32$,
though some effect is seen, in particular at $\beta = 1.0$,
for $L=8$. This lattice size dependence of fermion masses 
is consistent with the one of the correlation length of the scalar field, and
is very different from the size dependence of the fermion mass for 
asymptotic non-free cases\cite{aokinagai}.
These results
strongly indicate that
the chiral fermion spectrum observed on $L=16$ and 32 remains
in the infinite volume limit:
the chiral zero modes exist on the wall
in the infinite lattice volume of 2-dimensional physical
space-time.

\begin{figure}
\centerline{\epsfxsize12cm \epsfbox{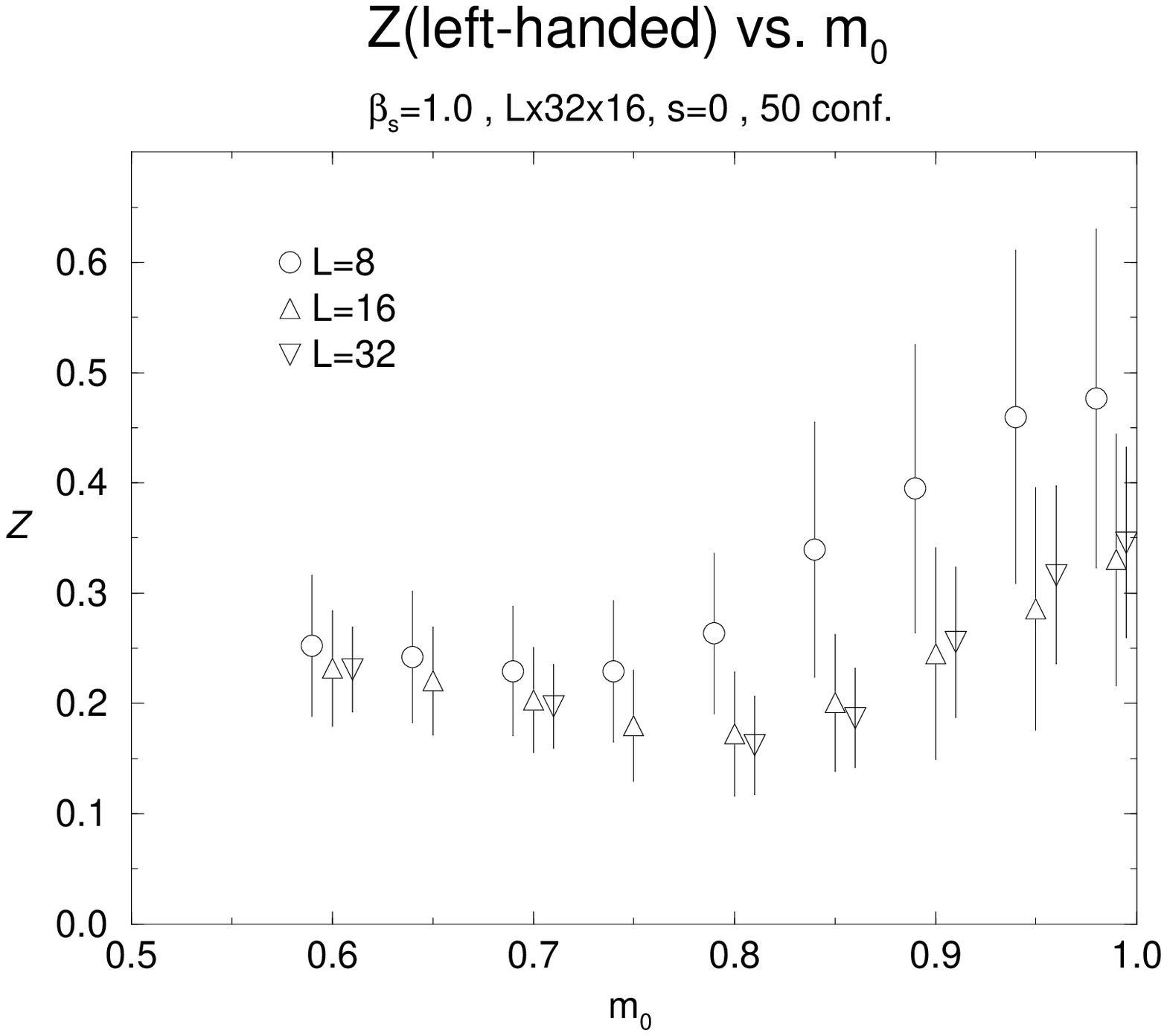}}
\centerline{\epsfxsize12cm \epsfbox{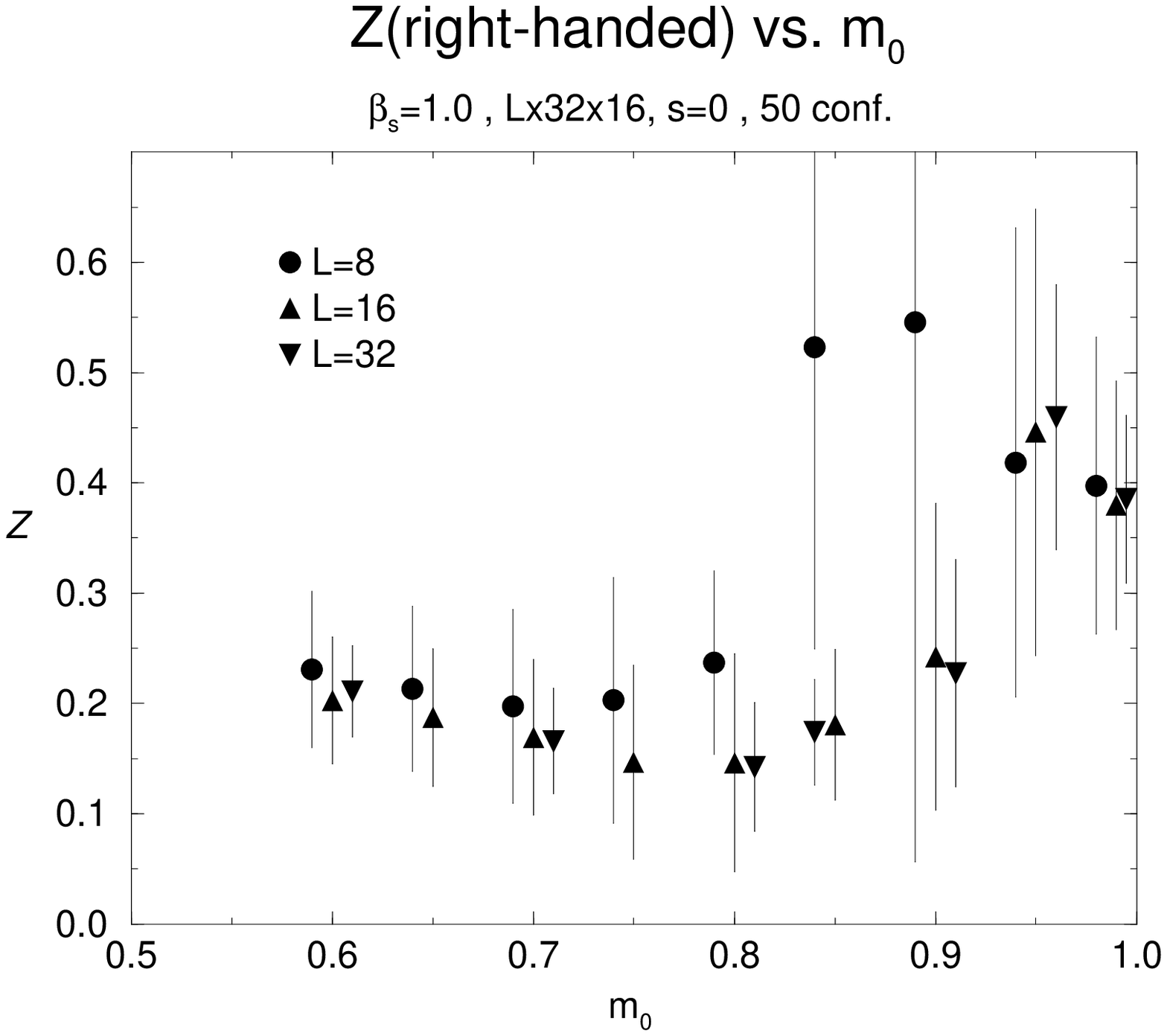}}
\caption{
(a) $z$(left-handed) vs. $m_0$ at $\beta_s=1.0$
(b) $z$(right-handed) vs. $m_0$ at $\beta_s=1.0$
on $L \times 32 \times 16$ lattice
with $L=$8(circles) , 16(up triangles) and 32(down triangles),
obtained form the fermion propagator on the domain wall at $s=0$.
}
\label{fig:zb1}
\end{figure}

\begin{figure}
\centerline{\epsfxsize12cm \epsfbox{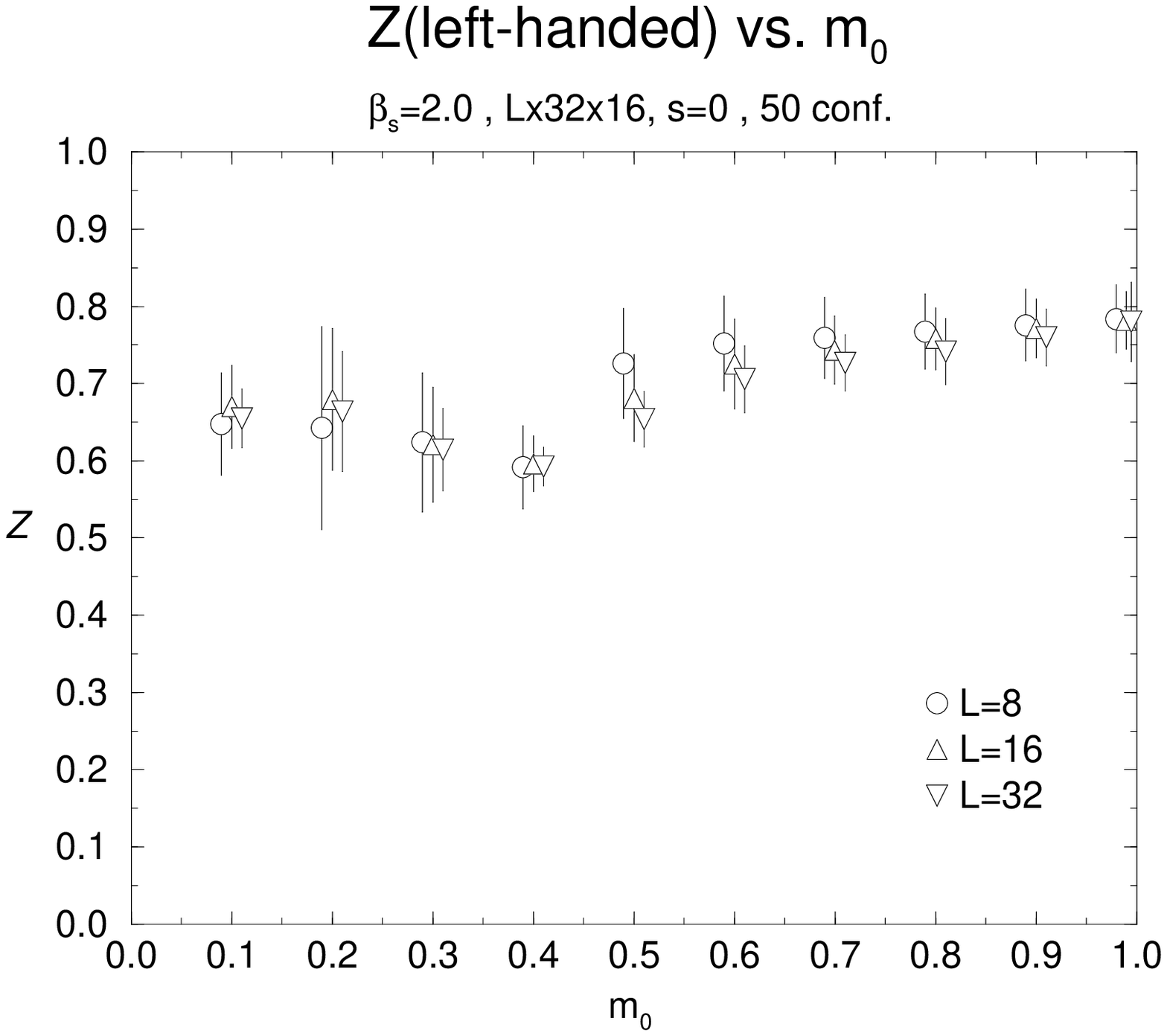}}
\centerline{\epsfxsize12cm \epsfbox{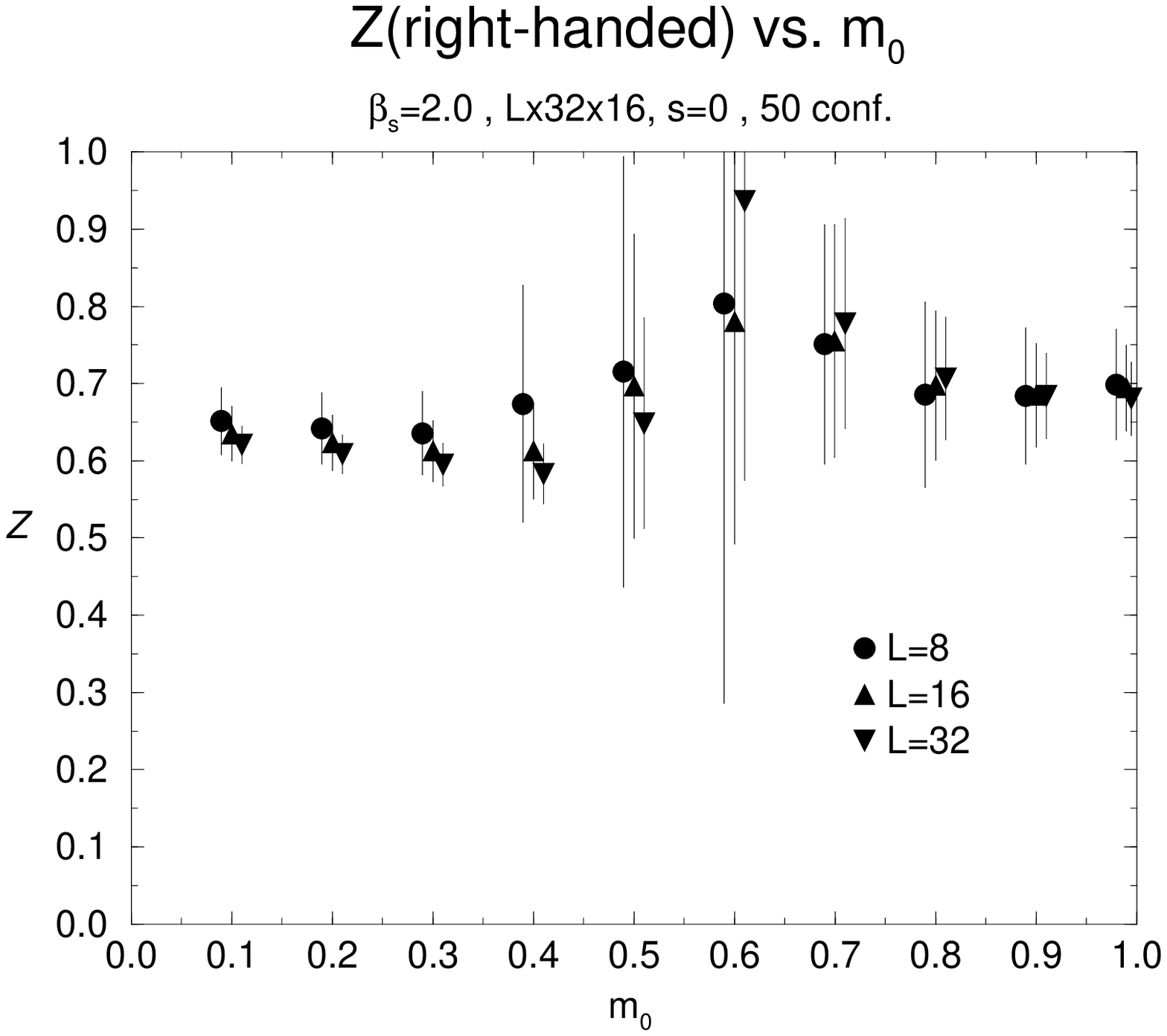}}
\caption{
(a) $z$(left-handed) vs. $m_0$ at $\beta_s=2.0$
(b) $z$(right-handed) vs. $m_0$ at $\beta_s=2.0$
on $L \times 32 \times 16$ lattice
with $L=$8(circles) , 16(up triangles) and 32(down triangles)
obtained form the fermion propagator on the domain wall at $s=0$.
}
\label{fig:zb2}
\end{figure}

In addition to the analysis on the fermion masses,
we carry out the mean-field analysis,
which has been shown to be successful to explain not only
the existence of the critical value of the domain wall mass,
but also the behavior of the fermion propagator\cite{aokinagai}.
We fit the measured propagator with the form of the mean-field
propagator obtained by replacing the link variable 
to some unknown but constant value $z$.
Note that non-zero value of $z$ means an existence of the chiral zero mode for 
the range that $ 1-z \le m_0 < 1$.
The extracted values of $z$'s are plotted as a function of $m_0$
in Figs.~\ref{fig:zb1} and \ref{fig:zb2}.
Within relatively large errors,
$z$'s are always non-zero at both $\beta_s$.
At $\beta_s =2$ $z$ depends very weakly on $m_0$
while it increases as $m_0$ approaches to 1 at $\beta_s =1$.
A similar $m_0$ dependence to the latter case has been also seen 
previously for asymptotic non-free cases\cite{aokinagai}.
Since $z$ should be constant on $m_0$ if the mean-field approximation
is exact, this $m_0$ dependence shows the magnitude of the error
for the mean-field approximation.
Again the finite size effect is small between $L=16$ and 32,
contrary to the case of the asymptotic non-free dynamics\cite{aokinagai},
and no indication that $z$ goes to zero as volume increases
is observed.
In conclusion
the mean-filed analysis also supports, at least at $\beta_s =2$,
the existence of the chiral zero mode on the wall in 
the infinite volume limit.

\section{Conclusions and Discussions}
\label{sec:conclusion}
In this paper we numerically investigate an existence of chiral zero modes
of the original domain-wall model in the presence of asymptotic-free dynamics
in 2(+1) dimensions
using the quenched simulation.
The result of the fermion mass on $16\times 32\times 16$ and
$32\times 32\times 16$ suggests that the chiral zero mode observed
on the domain-wall at $s=0$ will survive in the infinite volume limit.
The mean-field analysis of the fermion propagator
also supports this conclusion.

Although the positive indication is obtained for the construction of 
lattice chiral gauge theories via the domain-wall model with asymptotic-free
dynamics,
we have to understand a difference between an asymptotic-free and 
an asymptotic non-free cases more deeply.
It is usually thought that the vacuum expectation value of the scalar field
controls the existence of the chiral zero mode in the model: If non-zero
the zero mode exists, if not it does not.
In terms of the mean field analysis,
$z$ is supposed to correspond to the vacuum expectation value.
This is a crucial property for the failure of the domain-wall model in the 
symmetric phase, and
this seems true for an asymptotic non-free dynamics\cite{aokinagai}.
However the result of this paper shows that this does not hold anymore for
the asymptotic-free dynamics. Then, what controls the existence of zero modes ?
What corresponds to $z$ ? 
It is likely that some function of the correlation length does it.
Since the phase is unique for the SU(2)$\times$SU(2) non-linear $\sigma$
model in 2 dimensions, the function should be analytic in the correlation
length, so that it can vanish only on some points of $\beta_s$, but not
in the non-zero region of $\beta_s$. It is more likely that it vanishes
only at $\beta_s=0$. If this is true the zero mode exists for all $\beta_s$
except $\beta_s =0$. However since the scalar field near $\beta_s=0$ is 
very rough, the dynamics is very similar
to that in the symmetric phase of the asymptotic non-free model.
Therefore an existence of zero mode near $\beta_s=0$
may imply an existence of the zero mode also in the symmetric phase
of the asymptotic non-free model.
This seems inconsistent with the previous numerical results\cite{aokinagai}.
The one of solution to this puzzle is perhaps 
that the allowed range of the domain-wall
mass $m_0$ for the zero mode (: $m_c \le m_0 \le 1$ ) is too narrow to
be detected in the numerical simulation at the accuracy of 
Ref.~\cite{aokinagai}.
This point should be clarified,
in order to establish the existence of
the chiral zero mode of the asymptotic-free model without doubt.

Since scalar field theories with second derivative terms
are asymptotic non-free in 4 dimensions,
higher derivative terms have to be introduced to make them
asymptotic-free\cite{gdegree}. Since, in terms of gauge fields,
these higher derivative terms correspond to the gauge fixing
terms,
the solution to the rough gauge problem by the asymptotic-free dynamics
in 4 dimensions
is very similar to the recent proposal to the problem by the gauge fixing
for the U(1) theory\cite{BGS}.
The relation between the two should be understood.

If the rough gauge problem of the chiral zero mode in the domain-wall 
model can indeed be solved by an asymptotic-free dynamics, 
the rough gauge problems appeared in other models such 
as the Wilson-Yukawa model and the Waveguide model
are also resolved by the same dynamics.
Research in this direction is also necessary in order to
define lattice chiral gauge theories ultimately.
%
%
%
%
%----------------------------------------------------------------------
%
\section*{Acknowledgements}
Numerical calculations for the present work have been carried out
at Center for Computational Physics
and on VPP500/30 at Science Information Center,
both at University of Tsukuba. 
This work is supported 
in part by the Grants-in-Aid of the Ministry of Education
(Nos. 08640349, 09246206).
%
%
%

%
%
%--------------------------------------------------------------------
%
\newpage
%
%---------------------------------------------------------------------
%
%\clearpage
%

\begin{thebibliography}{l}
%
\bibitem{nielnino}
H. B. Nielsen and M. Ninomiya, 
Nucl.\ Phys.\ {\bf B185}, 20 (1981); 
erratum: {\it ibid.} {\bf B195}, 541 (1982); 
Nucl.\ Phys.\ {\bf B193}, 173 (1981); 
Phys.\ Lett.\ {\bf B105}, 219 (1981); \\
L. H. Karsten, 
Phys.\ Lett.\ {\bf B104}, 315 (1981).
%
\bibitem{kaplan}
D. B. Kaplan, 
Phys.\ Lett.\ {\bf B288}, 342 (1992).\\
For a recent review of domain-wall model, see, 
K. Jansen, Phys.\ Rep.\ {\bf 273} ,1 (1996) .
% 
\bibitem{wilyu}
W. Bock, A. K. De and J. Smit,
Nucl.\ Phys.\ {\bf B388}, 243 (1992);\\
M. F. L. Golterman, D. N. Petcher and E. Rivas,
Nucl.\ Phys.\ {\bf B377}, 405 (1992);\\
W. Bock, A. K. De, E. Focht and J. Smit,
Nucl.\ Phys.\ {\bf B401}, 481 (1993);\\
S. Aoki, H. Hirose and Y. Kikikawa
Int.\ J.\ Mod.\ Phys.\ A{\bf9}, 4129 (1994) .
%
\bibitem{aokinagai}
S. Aoki and K. Nagai, 
Phys.\ Rev.\ {\bf D53}, 5058 (1996);
Phys.\ Rev.\ {\bf D56}, 1121 (1997).
%
\bibitem{waveg}
M. F. L. Golterman, K. Jansen, D. N. Petcher, and J. C. Vink, 
Phys.\ Rev.\ {\bf D49}, 1606 (1994); \\
M. F. L. Golterman and Y. Shamir, 
Phys.\ Rev.\  {\bf D51}, 3026 (1995).
%
\bibitem{latchiral}
D. N. Petcher, Nucl.\ Phys.\ {\bf B} (Proc. Suppl.) {\bf 30}, 50 (1993); \\
Y. Shamir, Nucl.\ Phys.\ {\bf B} (Proc. Suppl.){\bf 47}, 212 (1996) .
%
\bibitem{gdegree}
Y. Kikukawa, KUNS-1445, hep-lat/9705024 (In two dimensions);
KUNS-1446, hep-lat/9707010 (In four dimensions).
%
\bibitem{overlap}
R. Narayanan and H. Neuberger,
Nucl.\ Phys.\ {\bf B412}, 574 (1994).
%
\bibitem{naraneu}
R. Narayanan and H. Neuberger, 
Phys.\ Lett.\ {\bf B302}, 62 (1993).
%
\bibitem{perturbation}
S. Aoki and H. Hirose
Phys.\ Rev.\ {\bf D49}, 2604 (1994);\\
S. Aoki and R. B. Levien,
Phys.\ Rev.\  {\bf D51}, 3790 (1995).
%
\bibitem{WZW}
Y. Kikukawa and S. Miyazaki, Prog.\ Theor.\ Phys.\ {\bf 96}, 
1189 (1996).
%
\bibitem{BGS}
Y. Shamir, TAUP-2306-95, hep-lat/9512019;
Nucl.\ Phys.\ {\bf B} (Proc. Suppl.) {\bf 53}, 664 (1997);\\
M. Golterman and Y. Shamir, Phys.\ Lett.\ {\bf B399}, 148 (1997);\\
W. Bock, M. Golterman and Y. Shamir, 
TAUP-2447-97, hep-lat/9708019;
HU-EP-97-64, hep-lat/9709113;
HU-EP-97-65, hep-lat/9709115; 
TAUP-2454-97, hep-lat/9709154.

%
\end{thebibliography}
\end{document}